\documentclass[aps,prl,reprint,superscriptaddress,twocolumn,groupedaddress,showpacs,amsmath,amsfonts]{revtex4}
\usepackage{graphicx}
\usepackage{bm}
\usepackage{color}

\def\Br{B_{\text{min}}/B_{\text{ECR}}}
\def\Bm{B_{\text{min}}}
\def\Be{B_{\text{ECR}}}

\begin {document}
\title {Observation of Poincar\'e--Andronov--Hopf bifurcation \\ in cyclotron maser emission from plasma magnetic trap}
\author{A.\,G.\,Shalashov}
\email{ags@appl.sci-nnov.ru}
\author{E.\,D.\,Gospodchikov}
\author{I.\,V.\,Izotov}
\author{D.\,A.\,Mansfeld}
\author{V.\,A.\,Skalyga}
\affiliation{Institute of Applied Physics, Russian Academy of Sciences (IAP RAS), 603950, 46 Ulyanova st., Nizhny Novgorod, Russia}
\author{O.\,Tarvainen}
\affiliation{Department of Physics, University of Jyvaskyla, PO Box 35 (YFL), 40500 Jyvaskyla, Finland}

\date{\today}

\begin{abstract}
We report the first experimental evidence of a controlled transition from the generation of periodic bursts of electromagnetic emission into continuous wave regime of a cyclotron maser formed in magnetically confined non-equilibrium plasmas. The kinetic cyclotron instability of the extraordinary wave of weakly inhomogeneous magnetized plasma is driven by the anisotropic electron population resulting from  electron cyclotron  plasma heating in MHD-stable minimum-B open magnetic  trap. 
\end{abstract}


\pacs{52.35.−g; 52.35.Qz; 52.55.Jd; 94.20.wj}

\maketitle

Electron cyclotron instabilities caused by resonant interaction between energetic electrons and  electromagnetic waves are typical phenomena for open plasma traps with magnetic mirrors. Studies of the cyclotron instabilities of non-equilibrium plasmas have led to the plasma cyclotron maser paradigm, which explains a rich class of phenomena of coherent radioemission from the Earth magnetosphere~\cite{trakh_books,Treumann_2006, speirs2014}, 
the Solar corona \cite{mss4,mss5, shal-trah,melrose,robinson,willes,wu}, other astrophysical objects \cite{Trigilio_2011_stars,mss1,mss2,mss3},
and laboratory magnetic traps \cite{VanCompernolle_2014,VanCompernolle_2015,VanCompernolle_2016,Bingham_2013,Ard_1966,Garner_1987,Garner_1990,
viktorov_rf2,viktorov_EPL,shal2017,Mansfeld_2007_JETP_maser,avod_whist,shalash_prl,shalash_ppcf}. 
Unlike vacuum electronic devices based on the electron cyclotron resonance \cite{Thumm}, stimulated emission from plasma traps is usually very far from being monochromatic. The obvious reason is the lack of electromagnetic mode selection by an external cavity typical for  masers and lasers. More neat reason is that the kinetic instabilities are driven by regions with free energy in the phase space of resonant particles, most commonly by the inverse population over Landau levels $\partial f/\partial v_\perp > 0$. 
In natural conditions, both space and laboratory, the distribution function of fast electrons is widely spread over the momentum space and inhomogeneous in the real space. The development of cyclotron instability under such conditions results in the generation of periodic, quasi-periodic or stochastic broadband pulses of emission. Each electromagnetic pulse is accompanied by pulsed precipitations of fast electrons from the trap as they lose transverse momentum and fall into the loss cone due to the interaction with waves. Owing to the sharp decrease in the free energy, the system falls under the instability threshold; after that, a comparatively slow preparation of the subsequent burst (accumulation of resonant particles) begins, and the process repeats \cite{trakh_books}. Indeed, starting from early works, the existence of oscillatory regimes  for constant strength of a particle source has been understood  as a general property of plasma systems with quasilinear relaxation  \cite{besp6,besp7,besp8,besp0}. 

Nevertheless, a  continuous wave (CW) generation is also possible when the system stays near the instability threshold, and the accumulation and emission phases are not separable, i.e.\ the number of high-energy particles delivered by a source is constantly equal to the number of precipitating particles. The steady-state emission of plasma cyclotron maser was extensively studied theoretically (see \cite{trakh_books} and references therein),
but it has  never been detected reliably in a laboratory because of the narrow region of plasma parameters where the regime exist.  

In this Letter, we present the first experimental evidence of the controlled transition between the burst to CW regimes of an electron cyclotron instability developing in the microwave range in an open magnetic trap and discuss the related physics. 
From a general point of view, this transition is related to the Poincar\'e--Andronov--Hopf bifurcation:  a stationary point attributed to CW generation becomes unstable through the birth of a stable limiting cycle. A similar transition is known for optic lasers with a nonlinear filter \cite{hanin}. In contrast to lasers, in which an active matter and a nonlinear absorber can be tuned independently, in the plasma maser both, the gain and dissipation, are governed by the same non-equilibrium electron distribution function. In this case, the transition to CW  regime requires fine tuning of the source of non-equilibrium electrons in phase space; in our experiment, we found an approach to such adjustment. 
%

Let us consider first a theoretical model that motivates us to perform the experiment. 
A self-consistent evolution of particles and waves may be described by the quasilinear theory, a perturbative approach that involves many overlapped wave-particle resonances as a basis for diffusive particle transport in a phase space \cite{QL00, QL001}. When the cyclotron instability evolves slowly in comparison to the bounce-oscillations of resonant electrons in an open trap  
and a narrow frequency spectrum of wave turbulence is assumed, a set of bounce-averaged quasilinear equations can be formulated as \cite{besp0,trakh_books}:
\def\k{\kappa}
\begin{equation}\label{eq1}\left\{\begin{aligned}
&\frac{\partial F}{\partial t} =E\;\frac{\partial }{\partial \k}\left(\!D\frac{\partial F}{\partial \k}\right)+J,
\quad \k=\frac{v_\perp}{v}\left(\!\frac{\Bm}B\!\right)^{1/2}\\
&\frac{\partial E}{\partial t} =E\int_0^\infty\!\!\!\int_{\k_c}^1 K\frac{\partial F}{\partial \k}\;d\k\; dv-\nu E
\end{aligned}\right.\end{equation}
where $F(t,\k,v)$ is the electron distribution function over the invariants of adiabatic motion, 
$\k_c=(\Bm/B_\text{max})^{1/2}$ is the loss-cone boundary, and $J(\k,v)$ is a stationary  source of non-equilibrium particles. The diffusion takes into account the scattering of electrons by unstable waves. The larger the average wave energy $E(t)$ in a magnetic-field tube, the faster is the electron diffusion into a loss-cone, which is the dominant mechanism of their loss. The wave energy, in turn, is determined from the averaged transport equation, in which the instability growth rate is proportional to $\partial F/\partial\k$, and $\nu$ stands for the wave dissipation due to damping in the background cold plasma and convective losses. Coefficients $D(\k,v)$ and $K(\k,v)$ are known smooth positive functions \cite{trakh_books}. For the extraordinary wave propagating along the magnetic field at the fundamental harmonic, one can neglect the dependence over $v$ in $D$ and consider $F(t,\k)$ as being integrated over the velocity modulus.


One can seek a solution of \eqref{eq1} as a series over the eigenmodes of the quasi-linear diffusion operator,
\def\l{\mu}
\begin{equation*}
F=\sum\nolimits_{n=1}^{\infty} f_n(t) \:\phi_n(\k), \quad J=\sum\nolimits_{n=1}^{\infty} j_n \:\phi_n(\k), 
\end{equation*}
then \eqref{eq1} are transformed into a set of balance equations 
\begin{equation}\label{eq2}
\frac{d{f}_n}{dt} =j_n-\l_n f_n E,\quad
\frac{d{E}}{dt} =\left(\sum\nolimits_{n=1}^{\infty} k_n f_n-\nu\right) E.
\end{equation}
Here the coefficients are defined as
\begin{equation*}
\frac{\partial }{\partial \k}D\frac{\partial \phi_n}{\partial \k}=-\l_n\phi_n, \quad
k_n=\int_0^\infty\!\!\!\int_{\k_c}^1 K\frac{\partial \phi_n}{\partial \k}\;d\k\; dv,
\end{equation*}
with the proper boundary conditions $\phi_n(\k_c)=0$ and $\phi_n'(1)=0$ being taken into account. Setting all time derivatives to zero, one can find a steady-state solution with non-zero wave energy,
\begin{equation}\label{eq3}
E^*=\nu^{-1}\sum\nolimits_{n=1}^{\infty} k_n j_n/\l_n, \quad f_n^*=j_n/(\l_n E^*).
\end{equation}
To study the stability of this state, let us note that eigenvalues $\l_n$ are rapidly growing with $n$. Therefore, one can simplify the analysis keeping only the lowest eigenmode $f_1$ in \eqref{eq2} and  assuming  all other modes adiabatically varying in time, $f_n=j_n/(\l_n E)$ for $n>1$~\cite{trakh_books}. Seeking perturbation  of the resulting second order equation in near the stationary solution \eqref{eq3} as $\delta E,\delta f_1\propto \exp(\lambda t)$, one finds the characteristic equation 
\begin{equation}\label{eq3a}\lambda^2+\left(\l_1 E^*+\sum\nolimits_{n=2}^{\infty}\frac{ k_n j_n}{\l_n E^*}\right)\lambda+\l_1\nu E^*=0.\end{equation}
The  boundary of stability corresponds to $\mathrm{Re}\lambda=0$, the steady-state is unstable when $\lambda$ is real or, equivalently, the term in the brackets is negative. For simplicity we may assume a weak source of non-equilibrium electrons, $E^*\ll \nu/\l_1$; then the instability criteria is  independent of the source power:
\begin{equation}\label{eq4}\sum\nolimits_{n=2}^{\infty} k_n j_n/\l_n<0.\end{equation}
Numerical integration of the complete set of balance equations \eqref{eq2} shows that the condition \eqref{eq4} predicts well a birth of a stable attractor (a limiting cycle) in all conditions relevant for experiments discussed hereafter.

We find that a key factor controlling the steady-state stability is the angular structure $\{j_n\}$ of the particle source. Let us assume that  hot electrons are accelerated by an external  wavefield with a frequency $\omega$ under \emph{off-center} electron cyclotron resonance (ECR) conditions in an open magnetic configuration. When the ECR surface is shifted outside the trap center and the plasma is rarefied for the heating wave ($k_{||}c/\omega<1$), the cyclotron interaction modifies the bounce-oscillations of a resonant electron along the magnetic field lines in such a way that the turning points move towards the ECR \cite{shalNF,NF37}. In the absence of other interactions, such electron would eventually turn exactly at the point of hot cyclotron resonance $\omega=\omega_{B}/\gamma$, where $\gamma=(1-v^2/c^2)^{-1/2}$ accounts for the cyclotron frequency downshift due to relativistic mass dependence (the Doppler shift is absent since $v_{||}=0$ at the turning point ). One may assume that the source delivers electrons with the same pitch-angle, in our notation
$$J=J_0\delta(\k-\k_\text{t}), \;j_n=J_0\phi_n(\k_\text{t}),\; \k_\text{t}=\sqrt{\Bm/\gamma\Be}.$$
Thus,  one may change the instability condition \eqref{eq4} by varying $\Br$ in a laboratory experiment. 

To illustrate this point, let us consider a simplified case with $D,K=\text{const}$. Then $\l_n=\pi^2(n-\frac12)^2/(1-\k_c)^2$, $k_n=(-1)^n$, and $\phi_n=\sin\sqrt\l_n (\k-\k_c)$. Since $k_1 j_1>0$, the lowest mode is always destabilizing in \eqref{eq2}. If   all modes with $n>2$ are ignored, one finds that the instability condition \eqref{eq4} leads to $k_2 j_2<0$. In other words, \emph{a stationary generation of electromagnetic radiation is stable when both modes are destabilizing, and a limiting cycle is stable when the second mode acts as a non-linear absorber}. The boundary between the regimes may be defined from $j_2(\k_\text{t})=0$, that results in a universal value of the bifurcation magnetic field $\Br=\gamma\k_t^2=\frac49\gamma(1+\k_c/2)^2$. For the fixed ECR heating frequency, the stationary generation corresponds to a higher magnetic field  in comparison to the burst regime. 

\begin {figure*}[tb]
\includegraphics [width=150mm] {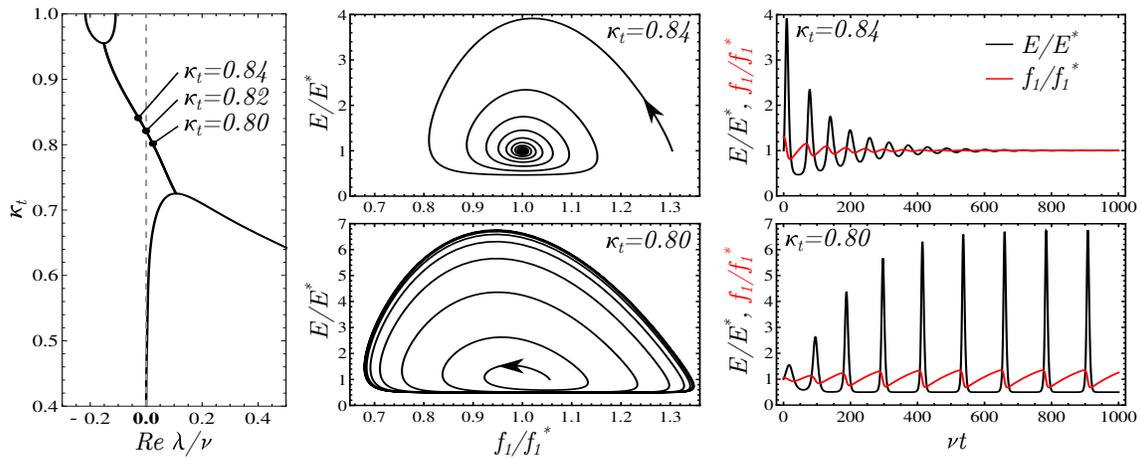}
\caption { \label{fig1}  Solution of the characteristic equation \eqref{eq3a}, phase trajectories and time-evolution for solutions of balance equations \eqref{eq2} in the limiting cycle (top) and continuous wave (bottom) regimes. The wave energy and number of particles in $\phi_1$-mode are normalized over its stationary values \eqref{eq3}. The parameters $J_0k_1/\nu^2=0.01$ and $\k_\text{c}=0.39$ are chosen to fit the experimental data; first 6 modes are taken into account in the numerical calculations.  The Hopf bifurcation occurs at $\k_\text{t}=0.82$.\\
}\end {figure*}

The same results are obtained for $D\propto\k$ and $K\propto\k^2$, that describe the extraordinary wave at the fundamental harmonic propagating at a small angle to the magnetic field (the case adequate to the experiment). Here  $\phi_n(\k)$ may be expressed analytically via Bessel functions, while $\l_n$ must be found numerically. 
Figure \ref{fig1} shows an example of such calculations for the parameters relevant to our experiment. 

\begin {figure}[tb]
\includegraphics [width=65mm] {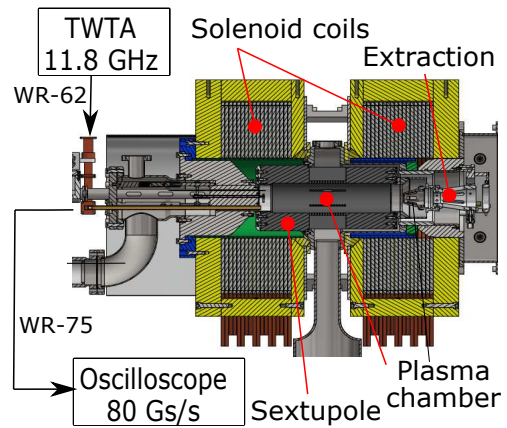}
\caption { \label{fig2} Schematic of the experimental setup. The mirror ratios are $B_{\text{max}}/\Bm=4.7-6.8$ and $2.4-3$, correspondingly, in the injection (left) and extraction  (right) mirrors. Microwave power (250\;W @ 11.8\;GHz) is launched 
through a WR-62 waveguide port, while the plasma microwave emission is measured through a WR-75 waveguide port, both incorporated into the injection iron plug. The transmission line to the oscilloscope (Keysight DSOV334A Infiniium V-Series, 80\;Gs/s sampling rate and 33\;GHz bandwidth) includes high voltage break, waveguide-to-coaxial transition, power limiter and tunable attenuator; the frequency response of the line is constant in the range of $8-15$\;GHz.  }
\end {figure}

The experiment was performed with the room-temperature A-ECR-U type electron cyclotron resonance ion source at JYFL accelerator laboratory \cite{ms1}. 
The setup is shown in Fig.~\ref{fig2}. The confining magnetic field was generated by two solenoid coils and a permanent sextupole magnet resulting in a minimum-B field configuration \cite{ms0}. 
A steady-state ECR plasma discharge was supported by microwaves at the frequency of 11.8\;GHz and the power range of $100-250$\;W provided by a traveling wave tube  (TWT) amplifier. 

The minimum value of the magnetic field $\Bm$ was achieved on the axis in between the solenoid coils. 
The ECR condition at the fundamental harmonic was satisfied on a closed (nearly ellipsoidal) surface with the constant magnetic field  $\Be = 0.42$\;T. The size of the ECR surface, characterized with the parameter $\Br$, was controlled by varying the solenoid coil currents. 
In the experiments, we run a continuous MHD-stable ECR discharge in oxygen at  pressure of $4-5\cdot 10^{−7}$\;mbar and tuned the control parameter in the range $ \Br =(0.75 - 0.99)\pm 1.5\%$.    
To reach this range, we reduced the heating frequency compared to the nominal value of 14 GHz.  

\begin {figure}[tb]
\includegraphics [width=85mm] {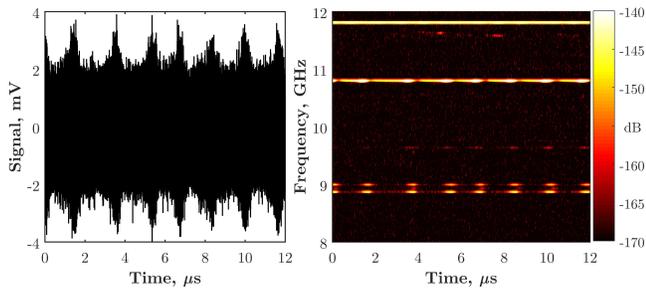}
\caption { \label{fig3} Recorded waveform of the plasma emission (left) and dynamic spectrogram in $8-12$\;GHz range (right). The dynamic spectra were calculated off-line by short-time Fourier transform with a Hamming window. Experiments were performed in oxygen plasma at pressure $4.5\cdot 10^{-7} \,$mbar,  TWT power  250\,W, and $\Br =0.935$.}
\end {figure}

Plasma microwave emission was detected with a high-bandwidth digital oscilloscope allowing direct recording of the waveforms of electromagnetic field emitted by the plasma with temporal resolution of 12.5 ps. More details on the diagnostic technique may be found in \cite{ms3,ms4}.
An example of the measured signal and its dynamic spectrogram in the frequency band of $8-12$\;GHz is presented in Fig.~\ref{fig3}. The emission spectrum consists of several  narrowband discrete lines with the line-width less than 30\;MHz. 

The experimental findings may be summarized as follows. No enhanced cyclotron (maser) emission from plasma is observed in low magnetic fields when $\Br<0.88$. A reproducible generation of quasi-periodic bursts near the fundamental and second electron cyclotron harmonics is observed at $\Br=0.88-0.93$. Similar regimes were reported previously \cite{ms4}. At higher values of the magnetic field, the CW generation near the fundamental harmonic is detected,  this occurs in the region $\Br=0.94-0.98$. 
With further increase of the magnetic field,  plasma heating becomes inefficient since the ECR absorption volume is small, and the cyclotron instability shows stochastic features.

\begin {figure}[t]
\includegraphics [width=80mm] {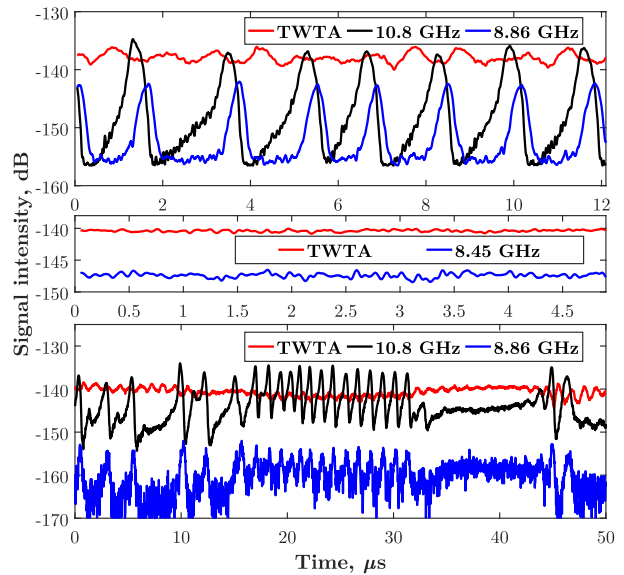}
\caption { \label{fig4} 
Intensity of microwave emission in $8-12$ GHz band referred to generation of periodic bursts (top), CW emission (middle), and spontaneous transition from quasi-periodic bursts to CW and back (bottom). Stray radiation of the heating TWT amplifier at 11.8 GHz is shown with a red line. Three regimes differ only by value of the external magnetic field: $\Br =0.935$ for the burst,  0.947 for the CW, and 0.98 for the transient regimes, correspondingly. Other experimental conditions are the same as in Fig.~\ref{fig3}.  }
\end {figure}

Typical patterns of the measured emission in the burst and CW regimes are presented in Fig.~\ref{fig4}. The top plot is related to the burst regime of the electron cyclotron instability. The microwave signal consists of series of wave packets with duration of $1\;\mu$s and repetition period of $2\;\mu$s. Depending on the experimental conditions, the  duration varies from 0.1 to $5\;\mu$s, simultaneously the period varies from $1\;\mu$s to 10\;ms.  
The corresponding dynamic spectrogram is shown in Fig.~\ref{fig3}---the frequencies of the most pronounced discrete lines are 10.8\;GHz, 9.0\;GHz, 8.86\;GHz. 
%
With the increase of the heating microwave power, the repetition period of pulses decreases, but the emission spectrum does not change significantly.

The middle  plot in Fig.~\ref{fig4} is related to the steady state regime. The frequency of CW plasma emission is 8.45 GHz and does not depend on the heating power. 

The bottom plot in Fig.~\ref{fig4} shows another interesting example of critical behavior not predicted by our theory. Here $\Br$ corresponds to the \emph{upper} boundary of the CW generation zone, at which the system randomly switches from generation of quasi-periodic series of pulses to continuous emission and back.

Although an absolute calibration of our detection system is complicated (as the radiation reabsorption and coupling efficiency are uncertain), we may estimate the power of the  microwave plasma emission as 800\;mW and 8\;mW, correspondingly, for the burst (within the pulse) and CW regimes. This essentially exceeds the power of the spontaneous electron cyclotron emission, that is estimated as 0.3\;mW (for $B\sim0.4$\;T, electron energy 200\;keV and density $10^9$\;cm$^{-3}$, plasma volume 100\;cm$^3$). 

Thus, all data suggest that the observed microwave emission is inherently related to the excitation of electromagnetic waves due to a kinetic cyclotron instability. The most unstable mode is apparently the slow extraordinary wave propagating quasi-longitudinally  to the external magnetic field and excited in the frequency range between the electron plasma and cyclotron frequencies \cite{ms4, ms7}. 
A significant part of the microwave power is measured at frequencies below the cyclotron frequency in the trap center, which indicates that the wave-particle interaction occurs at the relativistically down-shifted cyclotron resonance. To interact with the  CW radiation observed at 8.45\;GHz, the electrons must have the mean energy of 160\;keV. For this energy, our theory predicts the Hopf bifurcation at $\Br\!=\!\gamma\k_\text{t}^2\!=\!0.88$. To agree exactly with the experimental value 0.94, one must assume the electron energy of 200\;keV.
The  existence of fast electrons  in the considered energy range  has been confirmed independently from plasma bremsstrahlung and with a particle  detector for $1.5-400$\;keV~\cite{sk1} and  $5-250$\;keV ~\cite{sk2} electrons, respectively.  

The observed non-trivial dynamics caused by temporal modulation of the electron distribution function have been previously studied theoretically in the context of space cyclotron masers in planet magnetospheres and other astrophysical objects, and also have much in common with laser excitation mechanisms. 
Except being of fundamental interest, our results are important for applications, e.g.\ for the development of ECR ion sources. Particle ejections, which are inherent to the burst regime of cyclotron instability, cause oscillations of the plasma potential and  beam current accompanied with a significant decrease of the average ion charge~\cite{izo1,izo2}. 
The low-power CW regime would allow to avoid these non-desirable effects and improve the ion source performance.

\begin{acknowledgments}
This work was supported by the Academy of Finland under the Finnish Centre of Excellence Programme 2012-2017 (Project No. 213503) and mobility grants No. 296349  and No. 311237. The authors are grateful to Keysight Technologies Inc. for the technical support. 
\end{acknowledgments}


\begin {thebibliography} {99}

\bibitem{trakh_books}   P. A. Bespalov and V. Yu. Trakhtengerts, in \textit{Reviews of Plasma Physics}, vol.~10, edited by M. A. Leontovich (Consultants Bureau, New York, 1986);
V. Yu. Trakhtengerts and M. J. Rycroft   \textit{Whistler and Alfven mode cyclotron masers in space} (Cambridge University Press, New York, 2008)
\bibitem{Treumann_2006} R. A. Treumann. Astron. Astrophys. Rev., \textbf{{13}} {229} (2006).
\bibitem{speirs2014} D. C. Speirs, R. Bingham, R. A. Cairns, I. Vorgul, B. J. Kellett, A. D. R. Phelps, and K. Ronald, Phys. Rev. Lett. \textbf{113}, 155002 (2014)


\bibitem{melrose} D. B. Melrose and G. A. Dulk, Astrophys. J. \textbf{259}, 844 (1982)
\bibitem{wu}C. S. Wu, Space Sci. Rev. \textbf{41}, 215 (1985)
\bibitem{mss4} L. Vlahos, Solar Phys. \textbf{111}, 155 (1987)
\bibitem{robinson} P. A. Robinson, Sol. Phys. \textbf{134}, 299 (1991)
\bibitem{willes} A. J. Willes and P. A. Robinson, Astrophys. J. \textbf{467}, 465 (1996)
\bibitem{mss5} T. S. Bastian, A. O. Benz, D. E. Gary, Ann. Rev. Astron. Astrophys. \textbf{36}, 131 (1998)
\bibitem{shal-trah}  V. Yu. Trakhtengerts, A. G. Shalashov, Astron. Reports \textbf{43} 540 (1999)

\bibitem{mss3} C. Y. Ma, C. Y. Mao,  D. Y. Wang, X. J. Wu, Astrophys. Space Sci. \textbf{257}, 201  (1998)
\bibitem{mss1} R. Bingham,  R. A. Cairns,  B. J. Kellett, Astron. Astrophys. \textbf{370}, 1000  (2001)
\bibitem{mss2} M. C. Begelman,  R. E. Ergun,  M. J. Rees,  Astrophys. J. \textbf{625}, 51   (2005) 
\bibitem{Trigilio_2011_stars} C. Trigilio, P. Leto, G. Umana, C. S. Buemi and F. Leone, Astrophys. J. Lett. \textbf{739} L10 (2011)




\bibitem{Ard_1966} W. B. Ard,  R. A. Dandl,  R. F. Stetson, Phys. Fluids \textbf{9} (8) 1498 (1966)
\bibitem{Garner_1987}	Garner R.C., Mauel M.E., Hokin S.A. et al., Phys. Rev. Lett. \textbf{59} (16) 1821 (1987)
\bibitem{Garner_1990} R. C. Garner, M. E. Mauel, S. A. Hokin, R. S. Post and D. L. Smatlak, Phys. Fluids B-Plasma \textbf{2}, 242 (1990)
\bibitem{avod_whist}  A. V. Vodopyanov, S. V. Golubev, A. G. Demekhov, V. G. Zorin, D. A. Mansfeld,  S. V. Razin and V. Yu. Trakhtengerts, Plasma Phys. Rep. \textbf{31}, 927 (2005)
\bibitem{Mansfeld_2007_JETP_maser} A. V. Vodopyanov, S. V. Golubev, A. G. Demekhov, V. G. Zorin, D. A. Mansfeld, S. V. Razin and A. G. Shalashov, JETP \textbf{104}, 296 (2007)
\bibitem{shalash_prl} S. V. Golubev and A. G. Shalashov, Phys. Rev. Lett. \textbf{99}, 205002 (2007)
\bibitem{shalash_ppcf}  A. G. Shalashov, S. V. Golubev, E. D. Gospodchikov, D. A. Mansfeld and M. E. Viktorov, Plasma Phys. Control. Fusion \textbf{54}, 085023 (2012)
\bibitem{Bingham_2013} R. Bingham, D. C. Speirs, B. J. Kellett, I. Vorgul, S. L. McConville, R. A. Cairns, A. W. Cross, A. D. R. Phelps and K. Ronald, Space Sci. Rev. \textbf{178}, 695 (2013)
\bibitem{viktorov_rf2} M. E. Viktorov, S. V. Golubev, E. D. Gospodchikov, I. V. Izotov, D. A. Mansfeld and A. G. Shalashov, Radiophys. Quantum El. \textbf{56}, 216 (2013)
\bibitem{VanCompernolle_2014} B. Van Compernolle, J. Bortnik, P. Pribyl, W. Gekelman, M. Nakamoto, X. Tao, and R. M. Thorne, Phys. Rev. Lett. \textbf{112}, 145006 (2014)
\bibitem{VanCompernolle_2015}  B. Van Compernolle,  X. An,  J. Bortnik,  R. M. Thorne,  P. Pribyl, W. Gekelman, Phys. Rev. Lett. \textbf{114}, 245002 (2015)
\bibitem{viktorov_EPL} M. Viktorov, D. Mansfeld and S. Golubev, Eur. Phys. Lett. \textbf{109}, 65002 (2015)
\bibitem{VanCompernolle_2016} B. Van Compernolle, X. An, J. Bortnik, R. M. Thorne, P. Pribyl and W. Gekelman, Plasma Phys. Control. Fusion \textbf{59}, 014016 (2017)
\bibitem{shal2017}	A. G. Shalashov, M. E. Viktorov, D. A. Mansfeld, S. V. Golubev. Phys. Plasmas  \textbf{24} (3), 032111 (2017)


\bibitem{Thumm} M. Thumm, KIT Scientific Reports \textbf{7717}, 1 (2016)

\bibitem{besp6} P. A. Bespalov,  V. Yu. Trakhtengerts, Fizika Plasmy \textbf{2}, 396 (1976)
\bibitem{besp7}	H. L. Berk, T. D. Rognlien,  J.J.Stewart, Comments Plasma Phys. and Cont. Fusion \textbf{3}, 95 (1977)
\bibitem{besp8}	A. V. Gaponov-Grekhov, V. M.Glagolev, V. Yu. Trakhtengerts, JETP \textbf{80}, 2198 (1981)

\bibitem{besp0} P. A. Bespalov, Physica Scripta \textbf{T2/2}, 576 (1982)

\bibitem{hanin} Ya. I. Khanin. \textit{Fundamentals of Laser Dynamics.} (Cambridge International Science Publishing Ltd, 2006) p. 260

\bibitem{QL00} A. A. Vedenov, E. P. Velikov, R. Z. Sagdeev, Nucl. Fusion \textbf{1}, 82 (1961)
\bibitem{QL001} W. E. Drummond, D. Pines, Nucl. Fusion Suppl. \textbf{3}, 1049 (1962)

\bibitem{NF37} S. V. Golubev, V. E. Semenov, E. V. Suvorov, M. D. Tokman, Proc. Int. Conf. on Open Plasma Confinement Systems for Fusion (Novosibirsk, July 14--18, 1993) p. 261
\bibitem{shalNF} D. V. Yakovlev, A. G. Shalashov, E. D. Gospodchikov, A. L. Solomakhin, V. Ya. Savkin, P. A. Bagryansky, Nucl. Fusion \textbf{57}, 016033 (2017) 

\bibitem{ms1}  H. Koivisto,  P. Heikkinen, V. H\"anninen,  A. Lassila,  H. Leinonen,  V. Nieminen,  J. Pakarinen,  K. Ranttila, J. \"Arje,  E. Liukkonen,  Nucl. Instrum. Methods B \textbf{174}, 379 (2001)

\bibitem{ms0}  I. Izotov,  D. Mansfeld,  V. Skalyga,  V. Zorin,  T. Grahn,  T. Kalvas,  H. Koivisto,  J. Komppula,  P. Peura,   O. Tarvainen,  Phys. Plasmas \textbf{19}, 122501 (2012)

\bibitem{ms3} I. Izotov, T.Kalvas, H. Koivisto, R. Kronholm, D. Mansfeld, V. Skalyga, O. Tarvainen. Phys. Plasmas. Volume 24, Issue 4, 043515 (2017)
\bibitem{ms4} I. Izotov, O. Tarvainen, D. Mansfeld, V. Skalyga, H. Koivisto, T. Kalvas, J. Komppula, R. Kronholm, J. Laulainen, Plasma Sources Sci. Technol.  \textbf{24}, 045017 (2015)
\bibitem{ms7} D. Mansfeld, I. Izotov, V. Skalyga, O. Tarvainen, T. Kalvas, H. Koivisto, J. Komppula, R. Kronholm, J. Laulainen,  Plasma Phys. Control. Fusion  \textbf{58}, 045019 (2016)

\bibitem{sk1} T. Ropponen, O. Tarvainen, I. Izotov, J. Noland, V. Toivanen, G. Machicoane, D. Leitner, H. Koivisto, T. Kalvas, P. Peura, P. Jones, V. Skalyga and V. Zorin, 
Plasma Sources Sci. Technol. \textbf{20}, 055007 (2011)
\bibitem{sk2} I. Izotov, O. Tarvainen, V. Skalyga, D. Mansfeld, T. Kalvas, H. Koivisto, R. Kronholm, \textit{Measurement of the energy distribution of electrons escaping minimum-B ECR plasmas}. arXiv:1711.05562 [physics.plasm-ph]
\bibitem{izo1} O. Tarvainen, I. Izotov, D. Mansfeld, V. Skalyga, S. Golubev, T. Kalvas, H. Koivisto, J. Komppula, R. Kronholm, J. Laulainen and V. Toivanen, Plasma Sources Sci. Technol. \textbf{23}, 025020 (2014) 
\bibitem{izo2} O. Tarvainen, T. Kalvas, H. Koivisto, J. Komppula, R. Kronholm, J. Laulainen, I. Izotov, D. Mansfeld, V. Skalyga, V. Toivanen, G. Machicoane, Review of Scientific Instruments \textbf{87}, 02A703 (2016)


\end {thebibliography}
\end {document}